\documentclass[preprint,jcp,floats,amsmath,amssymb]{revtex4}

\usepackage{graphicx}% Include figure files
\usepackage{amsmath}
\usepackage{amssymb}
\usepackage[english]{babel}
\usepackage{MnSymbol}
\usepackage{color}
\usepackage{chemarr}

\begin{document}
\title{Validity conditions for moment closure approximations in stochastic chemical kinetics}
\author{David Schnoerr $^{1,2}$,  Guido Sanguinetti $^{2}$, Ramon Grima $^{1}$}
\affiliation{$^{1}$ School of Biological Sciences, University of Edinburgh, UK \\ $^{2}$ School of Informatics, University of Edinburgh, UK}

\begin{abstract}
Approximations based on moment-closure (MA) are commonly used to obtain estimates of the mean molecule numbers and of the variance of fluctuations in the number of molecules of chemical systems. The advantage of this approach is that it can be far less computationally expensive than exact stochastic simulations of the chemical master equation. Here we numerically study the conditions under which the MA equations yield results reflecting the true stochastic dynamics of the system. We show that for bistable and oscillatory chemical systems with deterministic initial conditions, the solution of the MA equations can be interpreted as a valid approximation to the true moments of the CME, only when the steady-state mean molecule numbers obtained from the chemical master equation fall within a certain finite range. The same validity criterion for monostable systems implies that the steady-state mean molecule numbers obtained from the chemical master equation must be above a certain threshold. For mean molecule numbers outside of this range of validity, the MA equations lead to either qualitatively wrong oscillatory dynamics or to unphysical predictions such as negative variances in the molecule numbers or multiple steady-state moments of the stationary distribution as the initial conditions are varied. Our results clarify the range of validity of the MA approach and show that pitfalls in the interpretation of the results can only be overcome through the systematic comparison of the solutions of the MA equations of a certain order with those of higher orders. 
\end{abstract}

\maketitle

\section{Introduction}

The chemical master equation (CME) is the well accepted mesoscopic description of chemical systems in well-mixed and dilute conditions \cite{Gillespie2007}. However, for most systems, analytic solutions are unknown. The stochastic simulation algorithm (SSA \cite{Gillespie1977}) is a popular Monte Carlo method for sampling from the probability distribution of the CME, but the SSA is computationally expensive and is tractable only for chemical reaction systems with a small number of reactions and for parameters such that the number of molecules is not too large.

For chemical systems solely composed of unimolecular reactions, the problems of the SSA are immaterial since the equations for the moments derived from the CME can be solved exactly \cite{McQuarrie1967}. However this is not the case for chemical systems with at least one bimolecular reaction which are the norm rather than the exception in nature. In this case the equations for the moments of the CME constitute an infinite hierarchy of coupled equations and hence they cannot generally be solved exactly. The roughest approximation to the problem involves solving the deterministic rate equations for the chemical system which leads to accurate estimates for the mean concentrations in the limit of large molecule numbers \cite{Kurtz1972}. Such a formalism is however inadequate when one is interested in estimating the size of the fluctuations, i.e., the variance in the fluctuations about the mean concentrations or when the goal consists in obtaining an approximate closed form solution for the probability distribution of the CME. These estimates are particularly important when the dynamics are strongly affected by the inherent stochasticity in the timing of chemical reaction events (intrinsic noise) such as the case when one or more chemical species are present in low abundances. 

Over the past few decades, several methods have been developed which provide a more accurate and complete picture than that obtained from deterministic rate equations. These methods involve solving a set of deterministic ordinary differential equations whose solution provides an approximation to the moments of the probability distribution of the CME. Two popular methods of this type are the linear-noise approximation \cite{vanKampen1961,ElfEhrenberg2003} and moment-closure approximations (MAs) \cite{GrimaJCP2012,Ferm2008,Ullah2009,Verghese2007,Ale2013}. The linear-noise approximation corresponds to the leading order term of the system-size expansion of the CME and hence the accuracy of its predictions and the range of its applicability has been deduced by considering the next to leading order terms of the expansion \cite{Grima2010,Thomas2012,Thomas2013}. In contrast MAs follow from an ad-hoc truncation of the infinite hierarchy of coupled moment equations of the CME and hence little is known about their range of validity and the accuracy of the moment estimates that they provide. Error estimates for the moments of the distribution of the CME of monostable chemical systems in the limit of large molecule numbers have been derived by Grima \cite{GrimaJCP2012}. A more fundamental question which has not been studied to-date is: when can we trust MAs to lead to physically meaningful estimates? i.e., positive real mean concentrations and positive real even central moments of the fluctuations in molecule numbers (note that by the latter we mean all moments of the type $\langle \prod_i (x_i - \langle x_i \rangle)^{k_i} \rangle$ such that $k_i$ is even for all $i$; this convention will be used throughout the paper). 

In this paper we report on a numerical study of a class of MAs which seeks to answer the aforementioned question. The paper is organised as follows. In Section II we provide an introduction to the mathematical framework of MAs, specify the class of MA methods which we will be concerned with and define a set of criteria which guarantee physical admissibility of the solution of the MA equations. In Section III, we numerically investigate the properties of the MA equations for four chemical reaction systems which are representative of deterministically monostable, bistable and oscillatory systems; we show that the MA equations lead to physically meaningful solutions only above a certain critical molecule number for deterministic monostable systems and only in a finite range of molecule numbers for deterministic bistable and oscillatory systems. We conclude in Section IV.

%%%%%%%%%%%%%%%%%%%%%%%%%%%%%%
%%%%%%%%%%%%%%%%%%%%%%%%%%%%%%
%%%%%%%%%%%%%%%%%%%%%%%%%%%%%%
\section{Moment closure approximations: definitions and criteria for physical validity}\label{sec_ma}

\subsection{Background}

Consider a chemical system involving species $X_i$ ($i=1,\ldots, N$) interacting via $R$ chemical reactions: 
\begin{align} 
\label{eq1}
  \sum_{i=1}^N  s_{ij} X_i  \xrightarrow{\quad c_j \quad} \sum_{i=1}^N r_{ij} X_i, 
   \quad j = 1, \ldots,  R.
\end{align}
Here, $c_j$ is the rate constant of reaction $j$. The stochiometric matrix is defined as $S_{ij}  =  r_{ij} - s_{ij}$. Under well-mixed and dilute conditions the system can be described by the joint probability distribution at time $t$,  $P(\mathbf{n},t)$, where $\mathbf{n}=(n_1, \ldots, n_N)$ is the state vector of the system and $n_i$ is the number of molecules of species $X_i$. Its time evolution is governed by the CME \cite{Gillespie2007}:
\begin{align}\label{cme}
  \partial_t P(\mathbf{n},t) 
  & = 
    \sum_{r=1}^R f_r (\mathbf{n} - \mathbf{S}_r) P(\mathbf{n} - \mathbf{S}_r, t) 
    - \sum_{r=1}^R f_r (\mathbf{n}) P(\mathbf{n}, t),
\end{align}
where $\mathbf{S}_r$ is the $r$th column vector of $S$ and $f_r(\mathbf{n})$ is the propensity function of reaction $r$ defined as \cite{vanKampen}:
\begin{equation}
f_r(\mathbf{n}) = c_r \Omega \prod_{k=1}^N \frac{n_k!}{(n_k - s_{kj})! \Omega^{s_{kj}}},
\end{equation}
and $\Omega$ is the volume of the system. To obtain the time evolution equation for the moment $\langle n_i \ldots n_l \rangle$ we multiply Eq.~\eqref{cme} by $n_i \ldots n_l$ and sum over all molecule numbers:
\begin{align}
  \partial_t \langle n_i \ldots n_l \rangle
  & = 
    \sum_{n_1, \ldots, n_N =0}^{\infty} n_i \ldots n_l \Bigg[ \sum_{r=1}^R f_r (\mathbf{n} - \mathbf{S}_r) P(\mathbf{n} - \mathbf{S}_r, t) 
    - \sum_{r=1}^R f_r (\mathbf{n}) P(\mathbf{n}, t) \Bigg], \\
  & = 
    \sum_{r=1}^R \langle (n_i+S_{ir})\ldots(n_l+S_{lr}) f_r (\mathbf{n}) \rangle
    - \sum_{r=1}^R \langle n_i \ldots n_l f_r (\mathbf{n}) \rangle.
\end{align}
For the first two moments one obtains:
\begin{align}
  \partial_t \langle n_i \rangle
  & = 
     \sum_{r=1}^R S_{ir} \langle f_r(\mathbf{n}) \rangle, \\
  \partial_t \langle n_i n_j \rangle
  & =
    \sum_{r=1}^R  \big[ S_{jr} \langle n_i f_r(\mathbf{n})  \rangle + S_{ir} \langle  f_r(\mathbf{n}) n_j \rangle
    + S_{ir} S_{jr} \langle  f_r(\mathbf{n})   \rangle \big].
\end{align}
For chemical systems composed of only unimolecular reactions, the moment equations are closed and can be solved explicitly. However for systems with at least one bimolecular reaction, this is not the case: the equation for a certain moment will depend on higher-order moments thus leading to an infinite hierarchy of equations which cannot be solved. The idea behind moment closure approximation involves the artificial truncation of this hierarchy at a certain order to obtain a finite set of equations that can be solved numerically. The truncation involves replacing all moments above a certain order by a function of the lower order moments. The latter function is ad-hoc and hence the accuracy of such approximations is not clear. A popular means of imposing the truncation involves setting the ($N+1$)th and higher-order cumulants to zero which leads to a closed set of equations for the first $N$ moments (see for example \cite{Ferm2008,Ullah2009,Verghese2007,Ale2013}). We will refer to this method as the $N$-moment approximation ($N$-MA) and exclusively focus on this class of moment closure approximations for the rest of this article. The conventional deterministic rate equations correspond to the 1MA, i.e, setting the variance to zero, and ignoring any factors of the form $\Omega^{-\alpha}$.

\subsection{Criteria for physical admissibility of the MA equations}

Here we formulate a set of criteria which guarantee physically meaningful predictions of MA approximations and which we will repeatedly use through the rest of this article. Given deterministic initial conditions, i.e., variance and all higher-order central moments are initially equal to zero, and provided the CME has a stationary solution for the probability distribution function, the MA equations should converge to a single steady-state in the limit of long time, and the trajectories should preserve a positive mean and even central moments in the molecule numbers for all times and for all initial conditions. 

Note that a \emph{unique steady-state solution of the MA equations} is generally expected for all systems independent of whether they are deterministically monostable, bistable or oscillatory. What we here mean by ``unique" (and throughout the rest of this article) is that given a fixed set of parameters, the time-dependent solution of the MA equations should converge in the limit of long times to the same fixed point for all possible initial conditions. Clearly this has to be the case since the stationary probability distribution of the CME is independent of initial conditions and hence the same steady-state moments must be reachable from all initial conditions. 

Note also that the \emph{long time solution of the MA equations should not show sustained oscillations for systems with time-independent rate constants}. This is since the moments of the CME in the limit of long time are always non-oscillatory (though the approach to the steady-state can be oscillatory) independent of whether the deterministic rate equations exhibit sustained oscillations or not. The explanation behind this phenomenon is that even if single trajectories of the SSA display sustained oscillations, independent trajectories get out of phase as time progresses and hence the ensemble-average over all the trajectories can only lead to non-oscillatory moments in the limit of long time. 

\section{Numerical analysis of the MA equations}

In this section, we show that the criteria set forth in Section II are not met by the MA equations for a number of chemical systems including some which are of biological relevance. We consider three types of systems: those whose rate equations have a single steady-state (deterministic monostable systems), those whose rate equations have two steady-states (deterministic bistable systems) and those whose are rate equations predict sustained oscillations (deterministic oscillatory systems). The chosen systems were selected since they are simple enough to study in depth while at the same time their behavior is representative of a large class of systems encountered in chemistry and biochemistry. Stochastic simulations for all systems except that in Section IIIA.1 where done using the software package iNA \cite{Thomas2012}.

\subsection{Deterministic monostable systems}\label{sec_mono}

\subsubsection{Bursty gene expression with no feedback}\label{subsec_dimerization}

We consider a model of bursty gene expression followed by a post-translational protein dimerisation reaction:
\begin{align}\label{gene_reactions}
  \varnothing \xrightarrow{\quad c_1 q_m \quad} m P, \quad m \in \mathbb{N}\, \backslash \{0 \}, \quad
  P + P \xrightarrow{\quad c_2 \quad} \varnothing,
\end{align}
where $P$ is the protein species, $c_2$ is the dimerisation rate constant, and $m$ is the burst size. Experimental \cite{Cai2006} and theoretical \cite{Thattai2001, Shahrezaei2008} evidence indicates a geometric distribution $q_m=p(1-p)^{m-1}$ with constant parameter $p$ ($p < 1$) as an appropriate model for bursting;  $c_1 q_m$ are then the rates at which bursts of size $m$ are created. The production step can be viewed as either an infinite number of input reactions, or equivalently as a single input reaction with input size $m$ being a random variable. Note that $1/p$ is equal to the mean burst size. Note also that in the limit of $p \approx 1$, the expression is non-bursty and the set of protein production reactions in (\ref{gene_reactions}) reduces to the single reaction $ \varnothing \xrightarrow{\quad c_1 \quad} P$.

We rescale time as $\tau=t c_2/ \Omega$ and define the dimensionless constant $k= \Omega^2 c_1/c_2$, where $\Omega$ is the volume of the system. It is easy to show that the rate equations for this system have a unique positive fixed point which is globally attractive for all $k$; exact stochastic simulations using the stochastic simulation algorithm (SSA \cite{Gillespie2007}) also show that the CME has a stationary solution for all values of $k$. 

Starting from the CME for reaction scheme (\ref{gene_reactions}), we can derive the time evolution equations for the first moment $\langle n \rangle$ and the second moment $\langle n^2 \rangle$ as described in Section \ref{sec_ma}:
\begin{figure*}[]
\centering
  \includegraphics[scale=0.5]{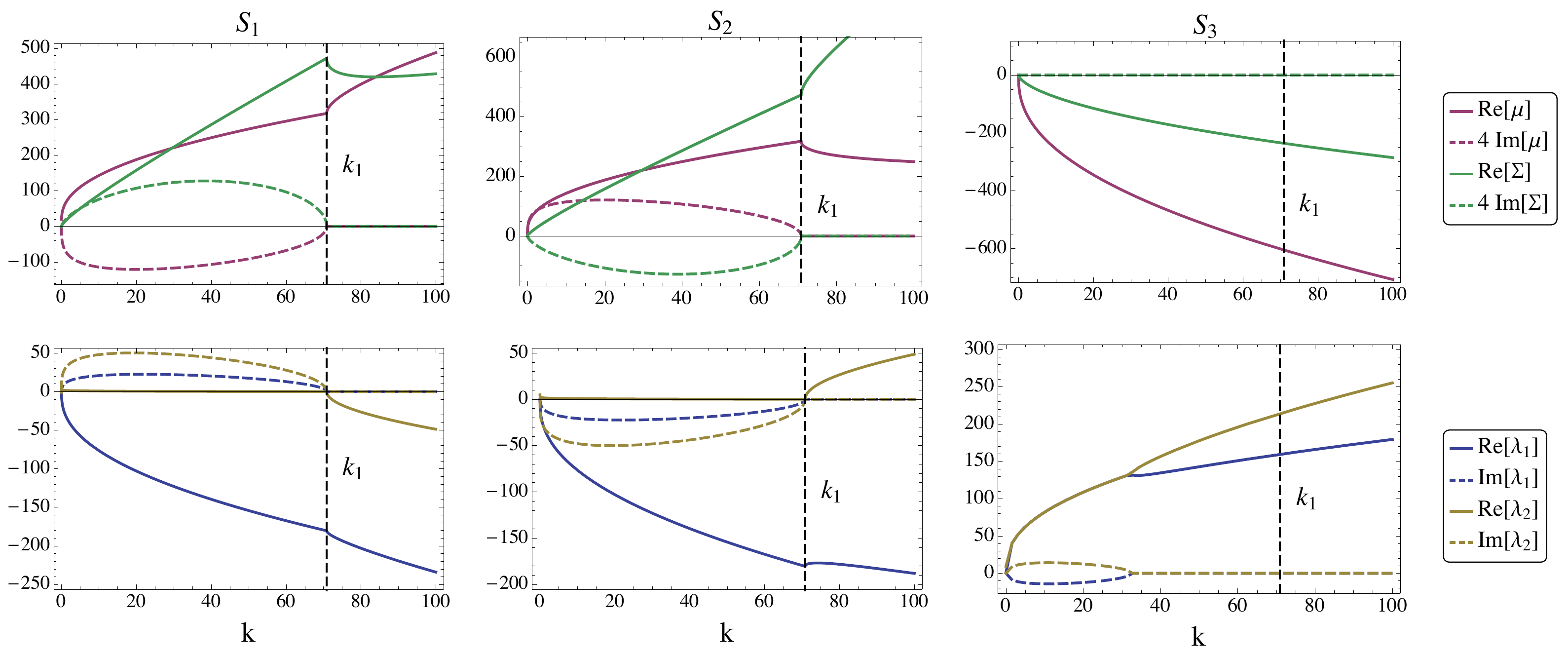}  
  \caption{Stability of the 2MA equations describing bursty gene expression (reaction scheme \eqref{gene_reactions}). The three plots in the upper panel show the real and imaginary parts of the mean protein number $\mu$ and the variance $\Sigma$ in protein number fluctuations of the three fixed points $S_i$ for the 2MA Eqs. \eqref{2ma_eq1}-\eqref{2ma_eq2} as a function of $k$ for a mean burst size of 20 (i.e., $p=1/20$).  The three plots in the lower panel show the associated real and imaginary parts of the eigenvalues of the Jacobian, $\lambda_1$ and $\lambda_2$, respectively. We find that $S_2$ and $S_3$ are always unstable, while $S_1$ is stable and the mean and variance are positive, if and only if $k > k_1$.}
  \label{figure_2ma_fixed_points}
\end{figure*}
\begin{align}
  \partial_{\tau} \langle n \rangle
  & = 
    -2 \langle n^2 \rangle + 2 \langle n \rangle + k \langle m \rangle, \\
  \partial_{\tau} \langle n^2 \rangle
  & =
    - 4 \langle n^3\rangle + 8 \langle n^2\rangle  - 4 \langle n \rangle +  2 \langle  n \rangle k \langle m \rangle  +  k \langle m^2 \rangle,
\end{align}
where $\langle m \rangle = 1/p$ and $\langle m^2 \rangle = (2-p)/p^2$ (these follow from the definition of $q_m$). These equations are then closed by using the 2MA (setting the third cumulant of $n$ to zero), leading to:
\begin{align}\label{2ma_eq1}
  \partial_{\tau} \mu
  & = 
    \langle m \rangle k + 2 \mu - 2 \mu^2 - 2 \Sigma, \\
\label{2ma_eq2}
  \partial_{\tau} \Sigma
  & = 
    \langle m^2 \rangle k + 4 \mu(\mu-1) + 8 (1-\mu)\Sigma,
\end{align}
where $\mu = \langle n \rangle$ and $\Sigma = \langle n^2\rangle - \langle n\rangle^2$ are the mean and variance in protein numbers. 

Setting the left hand side of Eqs.~(\ref{2ma_eq1})-(\ref{2ma_eq2}) to zero, and solving simultaneously, one finds that there are three possible solutions which we call $S_i=(\mu_i, \Sigma_i), i=1,\ldots,3$. Figure \ref{figure_2ma_fixed_points} shows the real and imaginary parts of the mean and variance of these three fixed points, as well as the real and imaginary parts of the associated eigenvalues of the Jacobian of Eqs.~(\ref{2ma_eq1})-(\ref{2ma_eq2}) for the case $p=1/20$ (this corresponds to a mean burst size of $20$ which has been measured experimentally for gene expression \cite{Cai2006}). By inspection of Figure \ref{figure_2ma_fixed_points}, we see that of the three possible steady-state solutions only $S_1$ is physically admissable; this is since it is the only steady-state solution which displays a positive mean and variance of molecule numbers and which is locally stable (negative real part of the eigenvalues of the Jacobian). However note that these properties only manifest for $k$ larger than a certain critical value $k=k_1 \approx 70$. This would lead one to surmise that the CME has a stationary solution only for $k$ greater than this critical value. However, as noted earlier this is not the case: the CME has a stationary solution for all values of $k$. These results taken together imply that \emph{the 2MA does not give a physically meaningful steady-state solution for all values of $k$}. 

\begin{figure}[t]
\centering
  \includegraphics[scale=0.24]{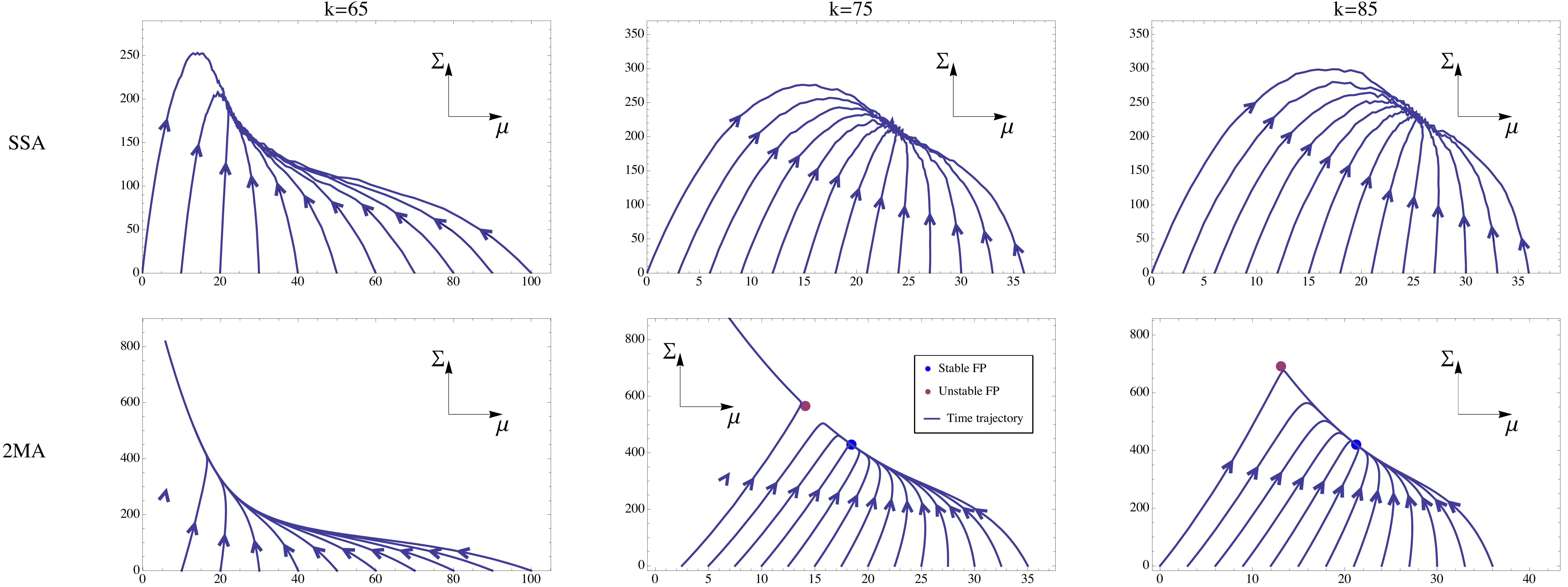}  
  \caption{Time trajectories of the 2MA equations and of the ensemble-averaged SSA in the $\mu$ - $\Sigma$ plane for different values of $k$ characterising bursty gene expression (reaction scheme \eqref{gene_reactions}). The initial conditions are deterministic. The blue and magenta dots show the stable and unstable fixed points (FPs) of the 2MA equations, respectively. While for $k=65$ there is no fixed point in the positive orthant and the time trajectories diverge for all initial conditions, for $k=75$ most initial conditions lead to trajectories converging to a unique stable fixed point. However, a small initial mean number of molecules still leads to divergence. Finally, for $k=85$ all initial conditions lead to trajectories converging to a unique fixed point with positive mean and variance. In contrast as SSA simulations show, the CME has a stationary solution for all values of $k$. The mean burst size is 20 ($p = 1/20$) as in Figure 1.}
  \label{figure_2ma_trajectories}
\end{figure}

Next we study the time-evolution leading to the steady-state. Figure \ref{figure_2ma_trajectories} shows the numerically integrated time trajectories for several initial conditions for three different $k$ values and for $p = 1/20$. We find that for $k < k_2 \approx 1.2 k_1 \approx 85$, \emph{some of the trajectories diverge as time goes to infinity} which is unphysical since the CME has a stable fixed point. This instability manifests for all deterministic initial conditions for $k < k_1$ and for initial conditions characterised by a small initial mean number of protein molecules for $k_1 \le k < k_2$. For $k \ge k_2$, however, the trajectories converge to the stable fixed point and are non-negative at all times. We verified this numerically for initial conditions up to $\mu(\tau=0)=10^6$. \emph{Hence in coincidence with the steady-state analysis above, we find that the 2MA only gives physically admissible solutions for a certain range of $k$.} Note that the requirement that the time trajectories of the moments are physically admissible at all times is harder to satisfy than the requirement that there exists a single physically admissible steady-state; this is since the critical value of $k$ above which the former is satisfied ($k_2$) is larger than the critical value of $k$ above which the steady-state criteria are satisfied ($k_1$). For the rest of this article we shall refer to the first requirement noted above as the time-dependent criterion and the second requirement as the steady-state criterion. 

\begin{figure}[t]
\centering
  \includegraphics[scale=0.3]{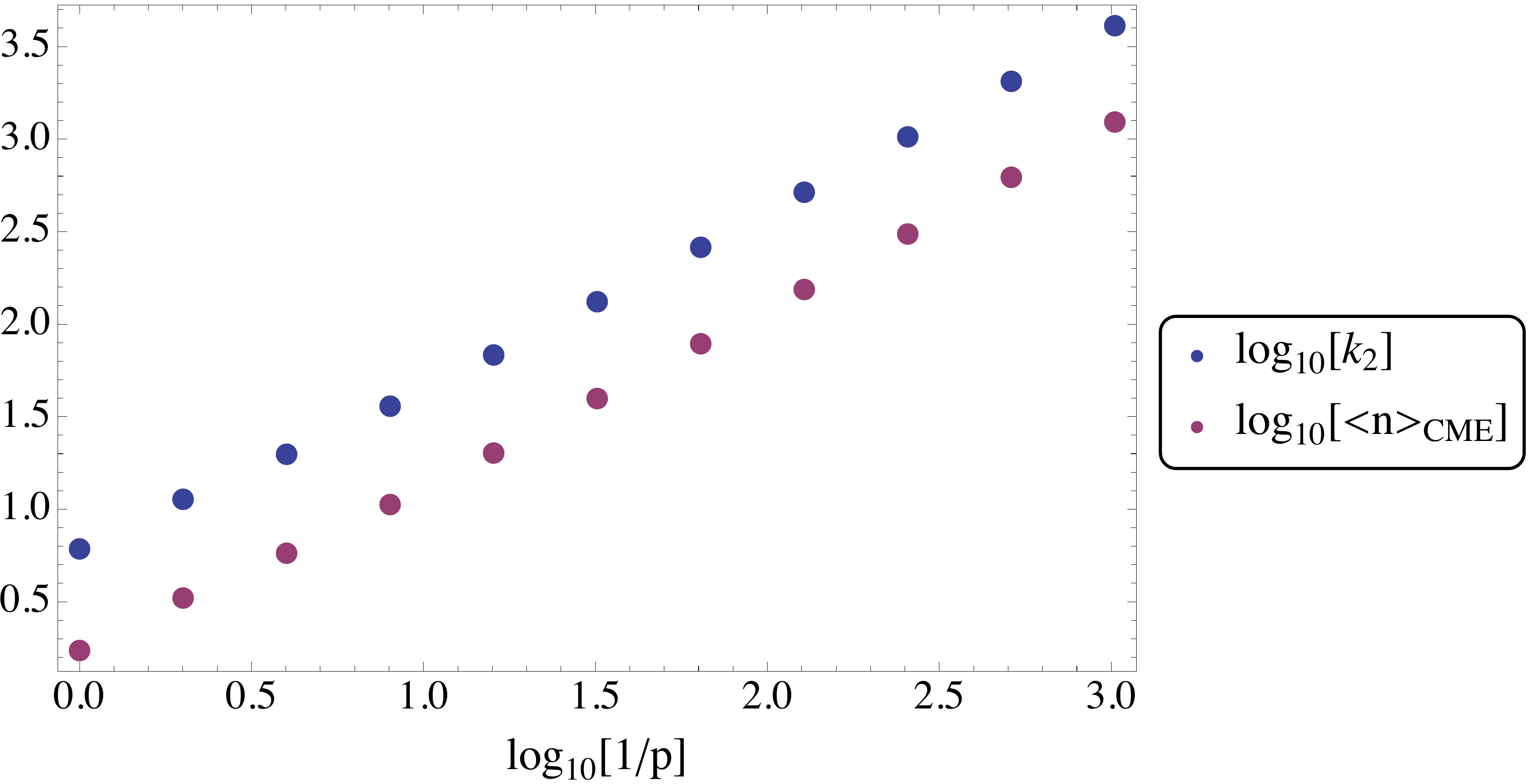}  
  \caption{Critical values ($k_2$, $\langle n \rangle_{\text{CME}}$) of the 2MA as a function of mean burst size $1/p$ for reaction scheme \eqref{gene_reactions}. Both curves are monotonically increasing with increasing $1/p$ implying that the critical molecule number above which the 2MA equations lead to physically meaningful results (both criteria in Section II B are satisfied) increases with burstiness in protein expression.}
  \label{figure_2ma_p_dependence}
\end{figure}

The analysis described above was specifically for the case of $p = 1/20$. Qualitatively similar results for the 2MA equations are found for all values of $p$, i.e, there exist a critical $p$-dependent values $k_1$ and $k_2$ such that for $k > k_1$ the steady-state criterion is satisfied and for $k > k_2$ the time-dependent criterion is satisfied; $k_2 \ge k_1$ such that the time-dependent criterion is generally the more difficult of the two criteria to satisfy. Figure \ref{figure_2ma_p_dependence} shows the p-dependence of the critical value $k_2$ as well as the p-dependence of the corresponding mean particle number $\langle n \rangle_{\text{CME}}$. Note that both $k_2$ and $\langle n \rangle_{\text{CME}}$ increase with $1/p$ implying that the larger the burstiness in protein expression, the larger is the critical molecule number above which the 2MA gives physically meaningful results. In particular for the case $p = 1/20$, we had earlier found that $k_2 = 85$ which corresponds to a mean steady-state protein number of $\langle n \rangle_{CME} \approx 25$, i.e., \emph{the 2MA equations for a mean protein burst size of 20 give physically meaningful results for the time-evolution of the system only when the number of protein molecules in steady-state exceeds 25}. It is well known that protein numbers per cell can be very small, even of the order of a few molecules and hence our results show that one must be careful in the use of the moment-closure approximation to understand cell level phenomena. 

\begin{figure}[t]
\centering
 \includegraphics[scale=0.24]{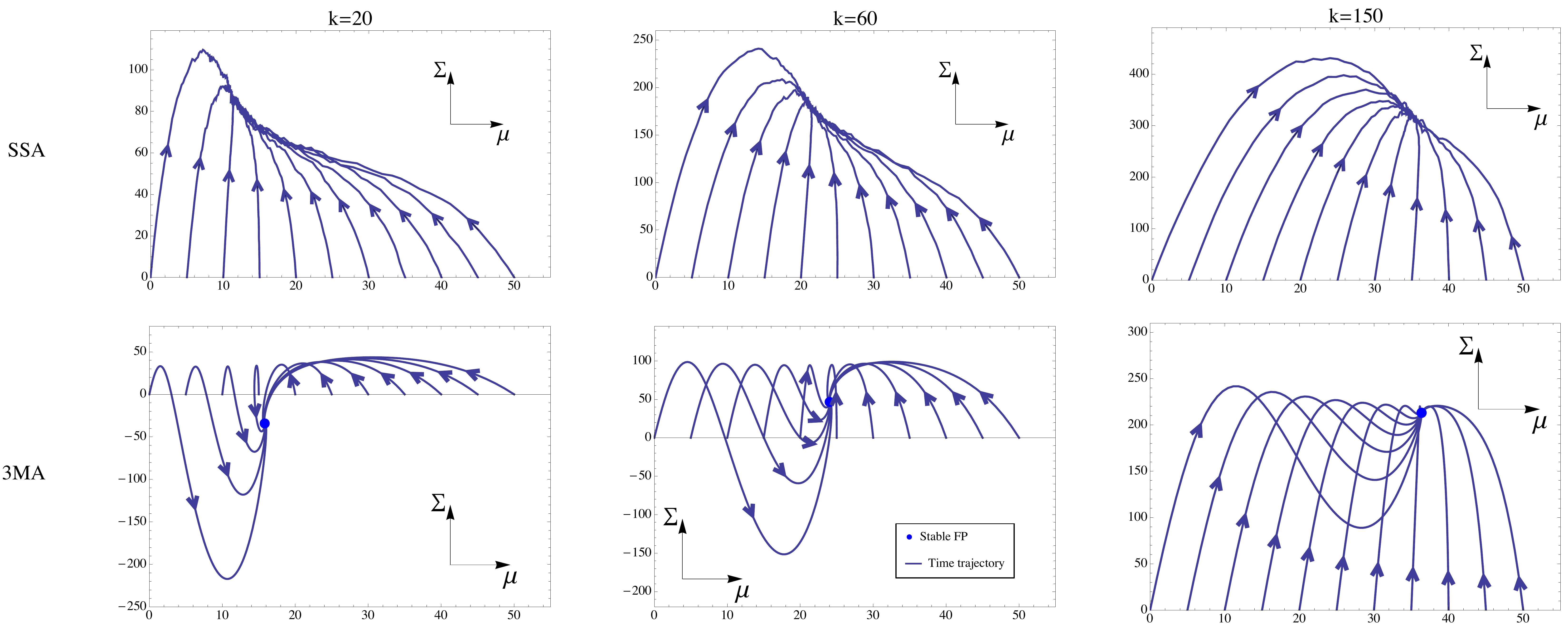}  
  \caption{Time trajectories of the 3MA equations and the ensemble-averaged SSA in the $\mu$ - $\Sigma$ plane for different values of $k$ characterising bursty gene expression (reaction scheme \eqref{gene_reactions}). For $k=20$ there is an unphysical stable fixed point of the MA equations with negative variance. For $k=60$ the stable fixed point has positive mean and variance but the variance becomes negative as the fixed point is approached. Finally, for $k=150$ all trajectories converge to a physically admissible steady-state and are physically meaningful (positive mean and variance at all times). In contrast the CME has a stationary solution for all values of $k$. The mean burst size is $20$ ($p = 1/20$). 
}
  \label{figure_3ma_p_trajectories}
\end{figure}

\begin{figure}[t]
\centering
  \includegraphics[scale=0.55]{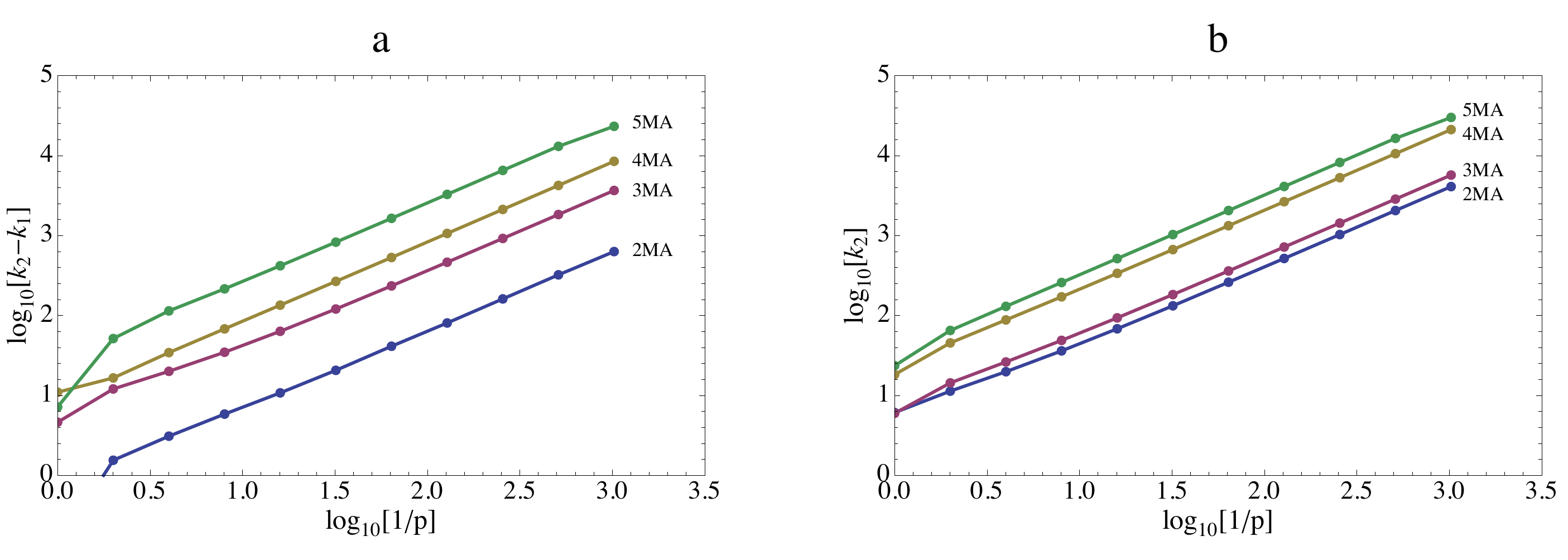} 
  \caption{Critical values $k_1$ and $k_2$ as a function of the mean burst size $1/p$ for the 2-5 MA equations describing bursty gene expression (reaction scheme \eqref{gene_reactions}). In panel (a) we show that $k_2 > k_1$ for all orders of the MA and for all $p$; hence the time-evolution criterion is satisfied over a smaller range of $k$ than the steady-state criterion. In panel (b) we show that $k_2$ increases monotonically with burstiness ($1/p$) and with the order of the MA. Hence the higher the order of the MA, and the larger the noise in gene expression, the smaller is the range of $k$ over which the MA equations give physically meaningful results for the time-evolution of the system.}
  \label{figure_mas_p_dependence}
\end{figure}

Similar results to the ones found for the 2MA are found for the higher-order moment closure approximations. In Figure \ref{figure_3ma_p_trajectories} we show the 3MA analog of the 2MA time-evolution analysis shown in Figure \ref{figure_2ma_trajectories}. The qualitative similarity between the two figures is evident: the time-evolution criteria are only satisfied for $k$ greater than a certain critical value ($\approx 100$) since for smaller values of $k$, we have a non-positive variance of molecule numbers as the steady-state is approached. We have also verified the same qualitative behaviour for the 4MA and 5MA (in addition to the mean and variance, we require the fourth central moment to be positive for these MA equations) which suggests that this behaviour exists for any order of the MA method. 

\begin{figure}[t]
\centering
  \includegraphics[scale=0.35]{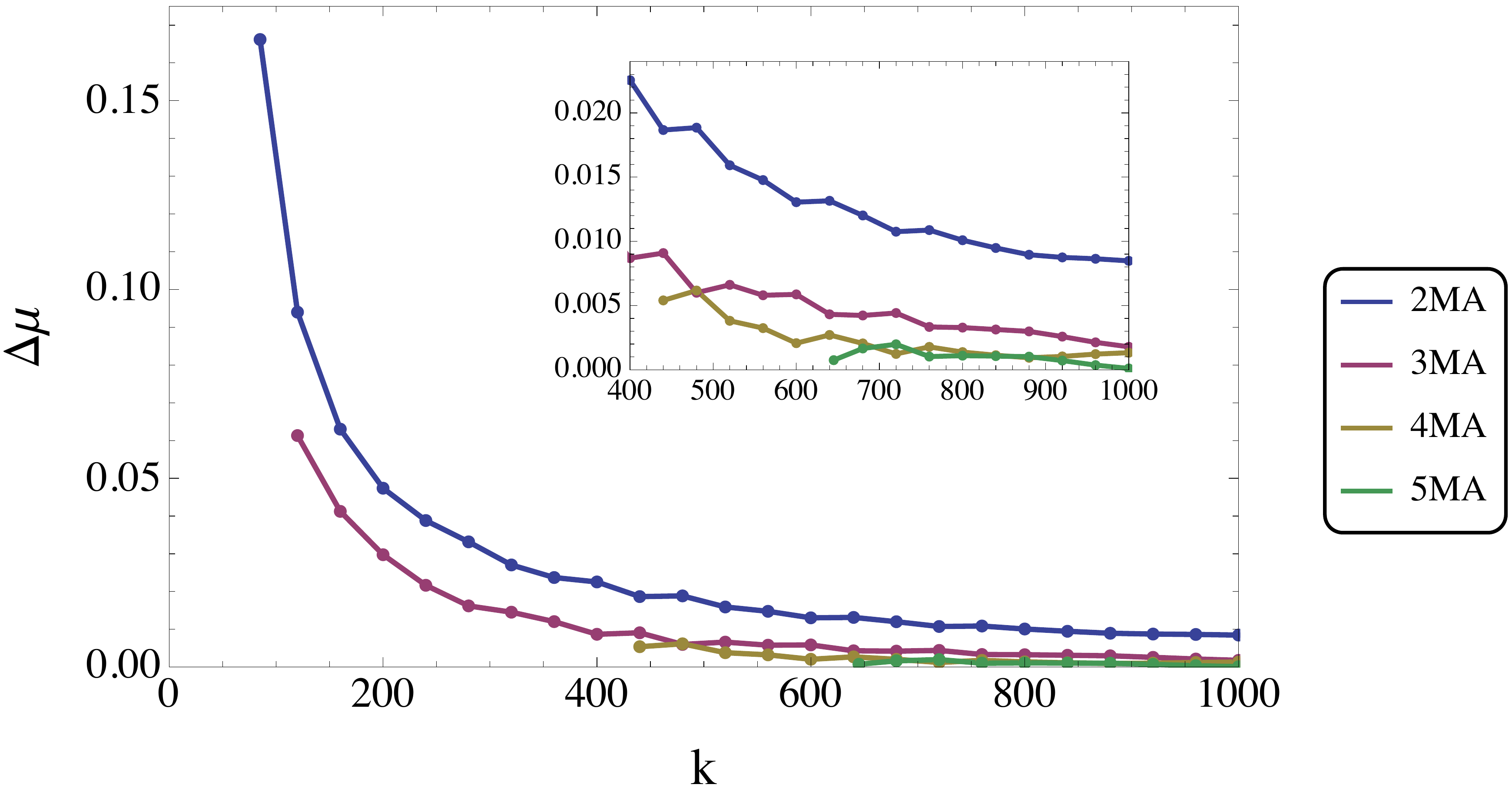} 
  \caption{Relative error in the mean number of molecules of the MA ($\Delta \mu$) as a function of the non-dimensional parameter $k$ and of the order of the MA. The error is computed as the absolute difference between the steady-state prediction of the first moment of the MA equations and the ensemble averaged SSA result for the mean number of molecules divided by the latter. The parameter $p$ is fixed to $1/20$. The accuracy of the MA increases with the order of the approximation while the range of validity of the MA equations decreases with the order (since the minimum value of $k$ at which the MA gives a physically meaningful result increases with order). }
  \label{accuracy_ma}
\end{figure}

Adopting the terminology that $k_1$ and $k_2$ are the values of $k$ above which the MA equations of any order satisfy the steady-state and time-dependent criteria, we show in Figure \ref{figure_mas_p_dependence} the dependence of these two values of $k$ on the mean burst size $1/p$ for all four MAs. Two observations can be made: (i) for all orders of the MA, $k_2 > k_1$ implying that the time-dependent criteria are the more difficult of the two to satisfy; this also implies that the mean molecule number associated with $k_2$ is the critical molecule number above which the MA equations give physically meaningful results. (ii) for a given $1/p$, the value of $k_2$ is monotonically increasing with increasing order of the MA, i.e., the range of molecule numbers over which one observes physically meaningful results decreases with increasing order of the MA. For $1/p = 20$, for example, we obtain the values $k_2 = 85, 116, 420, 643$ for the 2-5 MA equations respectively. Simulating the system using the SSA we find that these values correspond to mean protein numbers of $26, 31, 61$ and $75$, respectively. Finally in Figure \ref{accuracy_ma} we show a plot of the relative error in the mean number of molecules predicted by the MA as a function of $k$ and of the order of the MA; as expected accuracy increases with the order but the range of $k$ over which the approximation is valid decreases with increasing order. 

Summarising our analysis in this section shows that moment closure approximations for a gene circuit involving a bimolecular reaction give physically meaningful results (satisfy both criteria set forth in Section II B) only when the protein molecule numbers are above a certain critical threshold. This threshold increases with the mean burst size and closure order. 

%%%%%%%%%%%%%%%%%%%%%%%%%%%%%%
%%%%%%%%%%%%%%%%%%%%%%%%%%%%%%
\subsubsection{Gene expression with negative feedback}

We next consider the following gene regulatory network:
\begin{equation}\label{feedback_reactions}
  \begin{split}
  D_u & \xrightarrow{\quad c_1 \quad } D_u + X, 
  \quad D_u + X \xrightleftharpoons[\quad c_3 \quad]{\quad c_2 \quad } D_b,
    \quad   X  \xrightarrow{\quad c_4 \quad } \varnothing.
\end{split}
\end{equation}
A single gene in the unbound state $D_u$ expresses a protein $X$ which subsequently binds to the same gene and forms a non-expressive complex $D_b$. This is a negative feedback loop since the protein suppresses its own expression. Note that the mRNA is here not modelled explicitly for simplicity. It has been analytically shown that the rate equations possess a single steady-state solution and the CME has a stationary solution for the probability distribution function for all parameter values \cite{GrimaNewman2012}. Hence as for the previous example, we can check if the steady-state and time-dependent criteria are satisfied by the MA equations for this circuit. Note that this example unlike the previous one is two dimensional. 

%We rescale time as $\tau = t c_4$ and define the dimensionless parameters $k_1 = c_1/c_4, k_2 =c_2/c_4 \Omega, k_3 = c_3/c_4$. 
We fix the rate constants to $c_1/c_4 = 10, c_2/c_4 = 1, c_3/c_4 = 0.5$ and study the nature of the solutions of the MA equations as a function of the cell volume $\Omega$. Solving the 2-5MA equations in steady-state we find that they possess a unique stable fixed point with positive means and even central moments only above a certain critical volume $\Omega = \Omega_1$; in particular we find the values $\Omega_1 = 5.8, 16, 28, 39$, for the 2-5MA, respectively. Solving the 2-5MA equations as a function of time we find that for volumes less than the critical volumes stated above, the trajectories of the moments either diverge or converge to an unphysical fixed point characterised by a negative mean. In contrast for volumes larger than the critical volumes $\Omega_1$ all initial conditions tested give rise to convergent and physically meaningful time trajectories. These results imply that for this set of constants, the critical volumes above which the time-dependent criterion is satisfied are the same as the critical volumes above which the steady-state criterion is satisfied. 

From the exact solution of the CME for this circuit \cite{GrimaNewman2012}, we find that the mean protein numbers corresponding to the critical volumes of the 2-5MA are $3.3, 5.2, 6.3$ and $6.9$, respectively. Hence as for the previous example of bursty gene expression with no feedback, (i) the MA equations possess physically admissible solutions only when the steady-state molecule numbers predicted by the CME are above a certain minimum; (ii) these critical molecule numbers increase with the order of the MA equations. 

%%%%%%%%%%%%%%%%%%%%%%%%%%%%%%
%%%%%%%%%%%%%%%%%%%%%%%%%%%%%%
%%%%%%%%%%%%%%%%%%%%%%%%%%%%%%
\subsection{A deterministic bistable system}

Next, we consider a bistable reaction system that has been extensively studied in the literature, the Schl\"ogl system \cite{Schlogl1972} 
\begin{align}\label{schoegl_reactions}
  2X \xrightleftharpoons[c_2]{\quad c_1 \quad} 3X,
  \quad \varnothing \xrightleftharpoons[c_4]{\quad c_3 \quad} X.
\end{align}
The deterministic rate equations of the system \eqref{schoegl_reactions} have two steady-state solutions (bistability) in certain parameter regimes and one steady-state solution (monostability) in the rest of parameter space  \cite{Matheson1975}. The CME can also be solved exactly in steady state since the system is in detailed balance \cite{Matheson1975,Ebeling1979}. In what follows, we study the nature of the solutions of the MA equations for a parameter set in which the system is deterministically bistable and a parameter set for which it is monostable.  

%We rescale time as $\tau=t c_4$ and define the dimensionless parameters $k_1= c_1/c_4 \Omega, k_2= c_2/c_4 \Omega^2, k_3 = \Omega c_3 /c_4$. 

%%%%%%%%%%%%%%%%%%%%%%%%%%%%%%
%%%%%%%%%%%%%%%%%%%%%%%%%%%%%%
First, we consider the set of rate constants $c_1/c_4 = 2.6,  c_2/c_4 = 1.7,  c_3/c_4 = 0.1$ which we shall refer to as ``S1".  The deterministic rate equations have two positive stable fixed points for approximate concentrations, $0.16$ and $1.0$, respectively. In what follows we study the nature of the solutions of the MA equations for S1 as a function of the volume $\Omega$. Solving the time-dependent MA equations, we find that the requirement of convergent and physically meaningful time trajectories is only fulfilled above a critical volume. We find that the value of this critical volume increases with increasing closure order, ranging from $0.32$ for the 2MA to $5.6$ for the 5MA which corresponds to a range of mean molecule numbers (as calculated from the CME) equal to $\langle n \rangle \approx 0.02 - 1$. Hence as for the monostable circuits studied earlier, it is clear that the MA equations give valid meaningful results only above a certain critical molecule number. 

\begin{figure}[t]
\centering
  \includegraphics[scale=0.6]{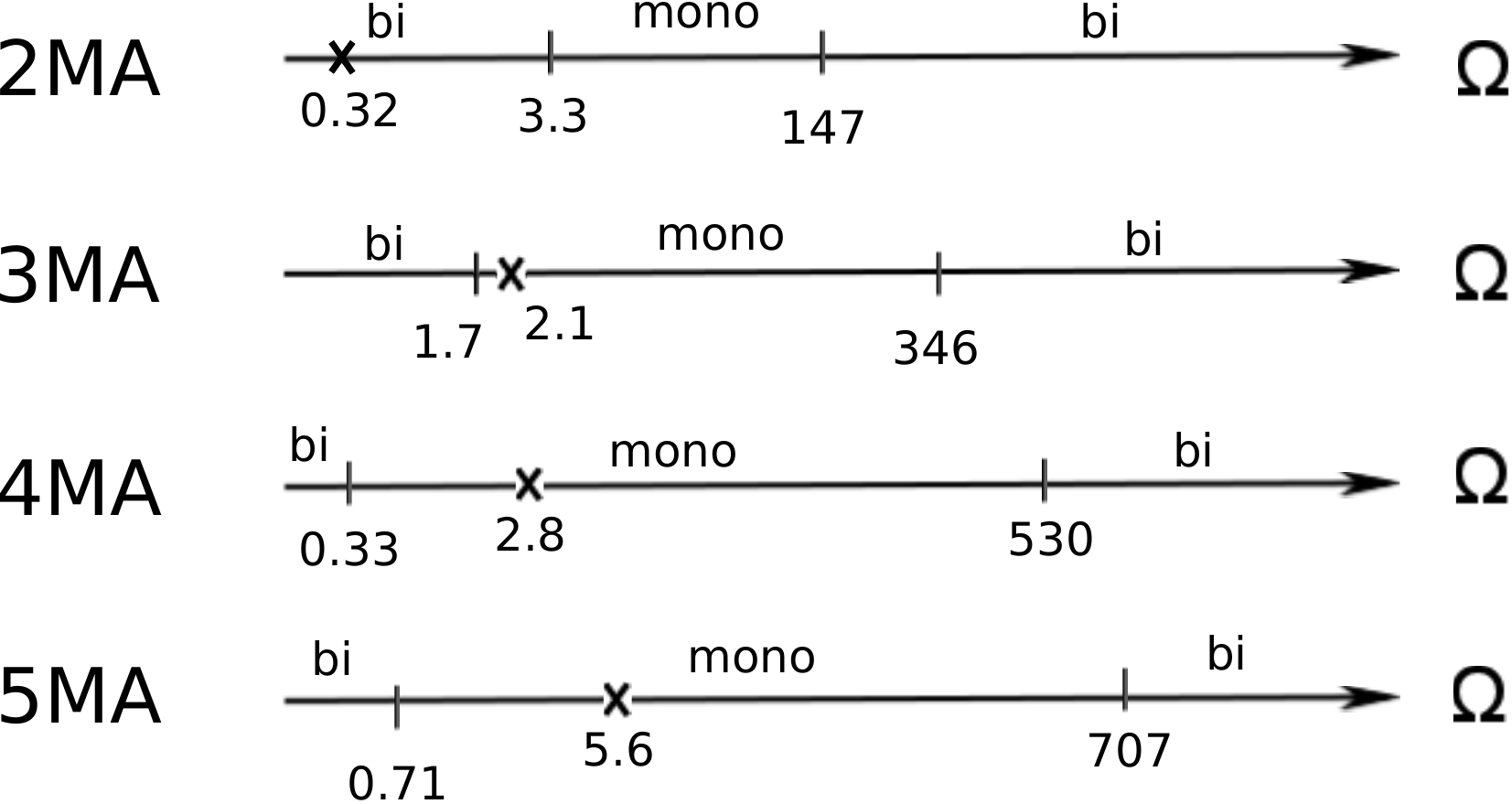}  
  \caption{Monostable and bistable regimes of the MA solution as a function of the volume $\Omega$ for the Schl\"ogl reaction scheme in \eqref{schoegl_reactions} with parameter set S1. The crosses denote the critical volumes above which all time trajectories for the mean and the even central moments are positive and convergent to a stable steady-state.}
  \label{figure_schloegl_arrow_bistable}
\end{figure}

\begin{figure}[t]
\centering
  \includegraphics[scale=0.38]{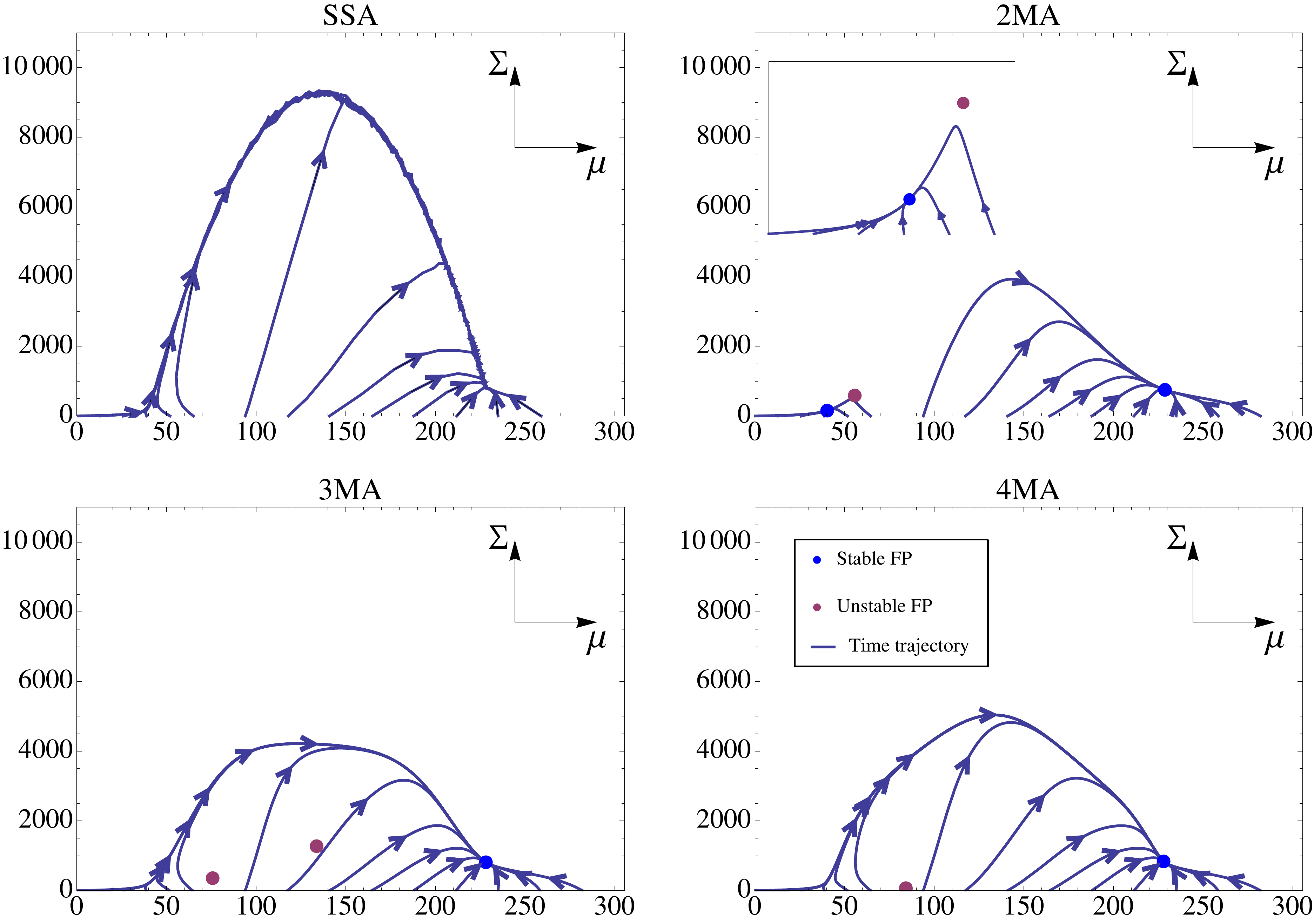}  
  \caption{Time trajectories of the SSA and 2-4MA equations for the Schl\"ogl reaction scheme in \eqref{schoegl_reactions} with parameter set S1 and volume $\Omega=235$. The inset for the 2MA shows how the time trajectories for small initial mean molecule numbers converge to one of two stable fixed points; time trajectories for large initial mean molecule numbers converge to the other stable fixed point. This bistable behaviour evident in the solution of the 2MA is ``washed away" as the order of the MA is increased and the time trajectories become similar to those of the SSA. This shows that the bistability in the 2MA solution at this volume should be considered as an artifact of the moment-closure approximation method.}
  \label{figure_schloegl_time}
\end{figure}

However there is a major difference between the solution of the MA equations for this system and the MA equations for the previous two monostable systems: the time-dependent MA equations converge to either one or two physically admissible steady-state values of the moments, depending on the volume. This bifurcation behaviour together with the critical molecule numbers described above are schematically summarised in Figure \ref{figure_schloegl_arrow_bistable}; the time-evolution of the 2-4MA equations are compared to the SSA in Figure \ref{figure_schloegl_time}. The non-uniqueness of a steady-state with positive mean and even central moments breaks the physical admissibility criteria set forth in Section II B since the moments of any probability distribution (independent of whether the probability distribution of the CME is unimodal or multimodal) should be single-valued. Hence on this basis one would conclude that the MA equations give physically admissible results only for volumes above the critical one and for which all trajectories converge to a single-valued solution. An inspection of Figure \ref{figure_schloegl_arrow_bistable} shows that this criterion implies that the 2-5MA equations lead to physically admissible solutions in the volume ranges $3.3-147$, $2.1 - 346$, $2.8 - 530$ and $5.6 - 707$ respectively. \emph{Thus for this deterministically bistable system, there are two critical volumes and not one as for monostable systems}. It is also the case that the size of the volume range increases with the order of the MA, this being mostly due to the fact that ceiling of this range (the volume at which the MA equations switch from one to two physically admissible steady-state solutions) increases with the order of the MA. The above results have also been verified for 8 other parameter sets for which the deterministic rate equations of the Schl\"ogl model are bistable (see Table I). Note that in a single case (first parameter set in Table I), there is no regime where the steady-state solution of the 2MA equations is physically meaningful since the solution is bistable for all volumes. 

The fact that the upper critical volume increases rapidly with the order of the MA is indeed an indirect verification that the regime in which there are two physically admissible steady-state MA solutions is an artifact of the approximation. The origin of the unphysical bifurcation is currently unclear although we have confirmed (see Appendix A) that it is not related to a sudden breakdown of the cumulant neglect assumption at the heart of the MA equations. 

\begin{table}
\begin{center}
  \begin{tabular}{|c|c|c|c|c|c|c|c|c|c|c|}
  \hline
  \multicolumn{3}{|c} {parameters} &  \multicolumn{2}{|c} {2MA} &  \multicolumn{2}{|c} {3MA} &  \multicolumn{2}{|c} {4MA} &  \multicolumn{2}{|c|} {5MA} \\ \hline
  $c_1/c_4$ & $c_2/c_4$ & $c_3/c_4$ & $\Omega_1$ & $\Omega_2$ & $~\Omega_1~$ & $\Omega_2$ & $~\Omega_1~$ & $\Omega_2$ & $~\Omega_1~$ & $\Omega_2$  \\ \hline \hline
 $ 1 $ & $ \frac{1}{8} $ & $ \frac{1}{8} $ & $ 0.0094 $ & $ 0 $ & $ 1.1 $ & $ 8.7 $ & $ 1.5 $ & $ 17. $ & $ 1.4 $ & $ 29000 $ \\ \hline$
 1 $ & $ \frac{1}{8} $ & $ \frac{1}{4} $ & $ 1.4 $ & $ 180 $ & $ 0.83 $ & $ 410 $ & $ 1.1 $ & $ 610 $ & $ 1.1 $ & $ 19000 $ \\ \hline $
 1 $ & $ \frac{1}{4} $ & $ \frac{1}{8} $ & $ 39. $ & $ 76. $ & $ 1.1 $ & $ 140 $ & $ 1.5 $ & $ 190 $ & $ 1.6 $ & $ 1.4\times
   10^6 $ \\ \hline $
 1 $ & $ \frac{1}{4} $ & $ \frac{1}{4} $ & $ 1.4 $ & $ 37. $ & $ 0.8 $ & $ 91. $ & $ 1.1 $ & $ 140 $ & $ 1.6 $ & $ 190 $ \\ \hline $
 2 $ & $ \frac{1}{8} $ & $ \frac{1}{8} $ & $ 2.6 $ & $ 4300 $ & $ 1.7 $ & $ 8600 $ & $ 2.1 $ & $ 72000 $ & $ 1.9 $ & $ 44000
   $ \\ \hline $
 2 $ & $ \frac{1}{4} $ & $ \frac{1}{8} $ & $ 2.6 $ & $ 1300 $ & $ 1.7 $ & $ 2800 $ & $ 2.1 $ & $ 4100 $ & $ 2.5 $ & $ 23000
   $ \\ \hline $
 2 $ & $ \frac{1}{2} $ & $ \frac{1}{8} $ & $ 2.6 $ & $ 370 $ & $ 1.7 $ & $ 810 $ & $ 2.1 $ & $ 1200 $ & $ 2.2 $ & $ 38000 $ \\ \hline $
 2 $ & $ 1 $ & $ \frac{1}{8} $ & $ 2.8 $ & $ 74. $ & $ 1.6 $ & $ 180 $ & $ 2.1 $ & $ 280 $ & $ 3.3 $ & $ 370 $ \\ \hline
  \end{tabular}
  \caption{The table shows the two volumes ($\Omega_1$ and $\Omega_2$) between which the MA equations have a physically meaningful solution for the Schl\"ogl reaction scheme in Eq.~\eqref{schoegl_reactions}. The data is generated for 8 distinct parameter sets. In the volume range $\Omega_1-\Omega_2$ the time trajectories of the MA are convergent to a unique steady-state and exhibit physically meaningful moments at all times (positive mean and even central moments). Above $\Omega_2$, the moments are not unique and below $\Omega_1$ the time trajectories do not converge or else there exist negative mean and even central moments at a point in time. We consider all combinations for the parameters $c_1/c_4$, $c_2/c_4$ and $c_3/c_4$ drawn from the set $\{ 2^{-3}, 2^{-2}, \ldots, 2^3\}$ for which the rate equations are bistable.}
  \label{table_schloegl}
  \end{center}
\end{table} 

We note that another interpretation of the phenomenon may at first appear possible. When the deterministic rate equations possess two steady-state solutions the probability distribution of the CME for large volumes is expected to be bimodal; similarly when the rate equations possess one steady-state then the distribution is unimodal for large volumes. Now if one sees the MA equations as a refinement of the conventional rate equations, in the sense that they are valid not only for large volumes but over a wider range of volumes, then one may postulate that the number of physically admissible steady-state solutions of the MA equations reflects the number of maxima of the probability distribution of the CME. As we now show this interpretation is not valid for the Schl\"ogl model. For the parameter set $S1$, the probability distribution of the CME is unimodal for volumes below $\Omega = 30$ and bimodal for larger volumes. In contrast the 2MA equations show two transitions from bimodal at low volumes to unimodal at intermediate volumes to bimodal at large volumes (see Figure (\ref{figure_schloegl_arrow_bistable})); furthermore the volumes at which the transition from unimodal to bimodal occurs ($\Omega = 147, 346, 530, 707$ for the 2-5MA respectively) bear no relationship to the actual transition volume $\Omega = 30$ obtained from the CME. Hence we arrive to the conclusion as before, namely that generally the bifurcations in the solutions of the MA equations are artificial in the sense that they are not indicative of any real transition in the number of maxima of the probability distribution of the CME. Preliminary analysis (see Appendix B) does however suggest that the two steady-state solutions of the MA equations contain information (position and width) on the two peaks of the bimodal distribution of the CME but not on the height of the two peaks; this information is partial in the sense that it cannot be used to reconstruct the probability distribution or to calculate any of its moments and furthermore the information about the individual peaks is only valid over a subset of the volumes over which the probability distribution of the CME is bimodal. \emph{Hence in line with our previous analysis, it can be concluded that it is only safe to trust information from the MA equations when the time-evolution of the trajectories is physically meaningful (positive mean and even central moments) and when they converge to a single steady-state.}

%%%%%%%%%%%%%%%%%%%%%%%%%%%%%%
%%%%%%%%%%%%%%%%%%%%%%%%%%%%%%

\begin{figure}[t]
\centering
  \includegraphics[scale=0.6]{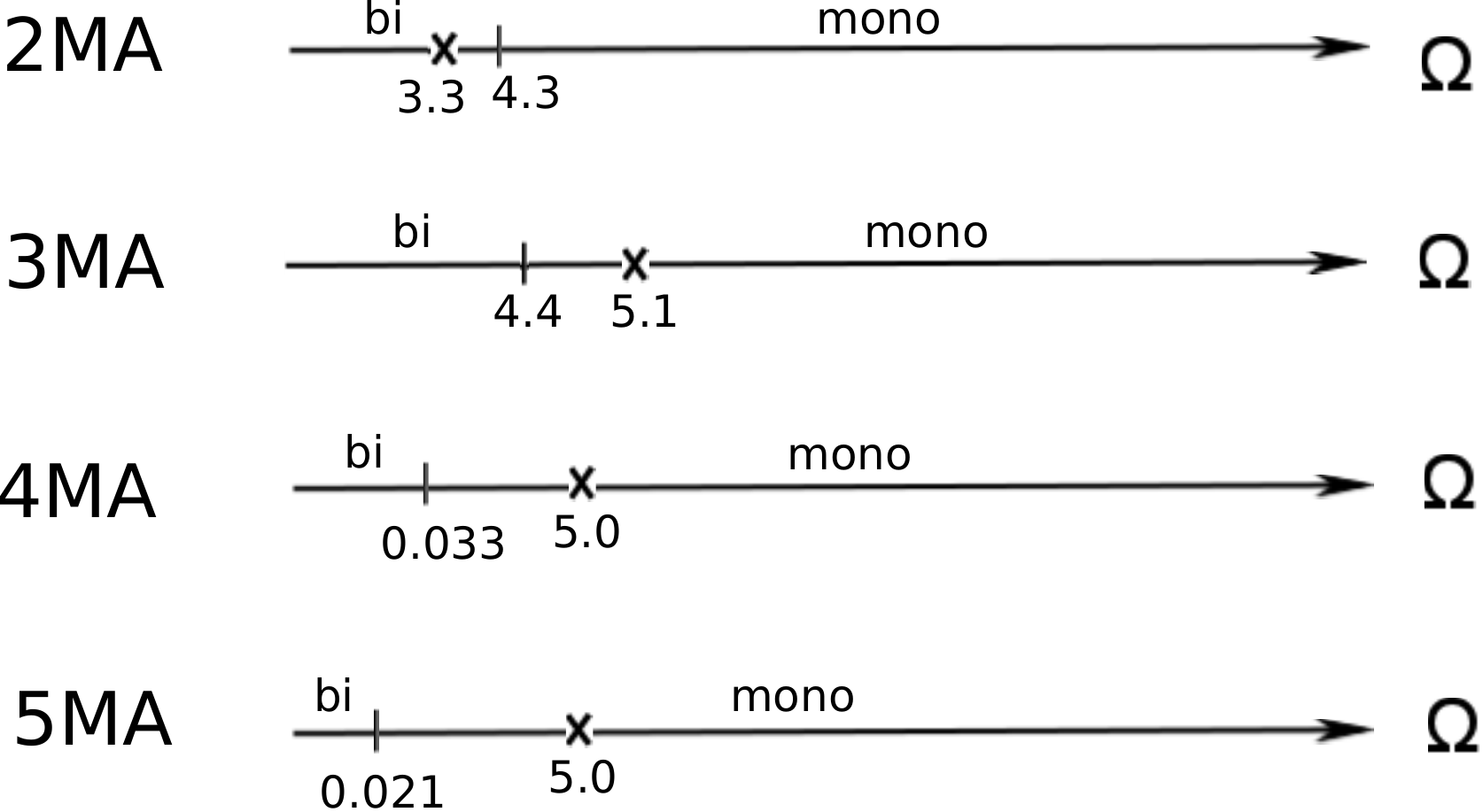}  
  \caption{Monostable and bistable regimes of the MA solution as a function of the volume $\Omega$ for the Schl\"ogl reaction scheme in \eqref{schoegl_reactions} with parameter set S2. The crosses denote the critical volumes above which all time trajectories for the mean and the even central moments are positive and convergent to a stable steady-state.}
  \label{figure_schloegl_arrow_uni}
\end{figure}

Next, we consider the set of rate constants $c_1/c_4 = 10,  c_2/c_4 = 0.1,  c_3/c_4 = 0.1$ which we shall refer to as ``S2".  The deterministic rate equations have now one positive stable fixed point and the probability distribution of the CME is unimodal for all volumes that we checked. As for the monostable circuits studied in Section III A, the MA equations only give physically meaningful results (time-dependent criterion is satisfied) above a certain critical volume. These together with the number of steady-state solutions of the MA equations are summarised in Fig. \ref{figure_schloegl_arrow_uni}. The critical volumes range from $\Omega \approx 3.3$ for the 2MA to $\Omega \approx 5$ for the 5MA and the associated mean molecule numbers (calculated from the CME) range from $\langle n \rangle \approx 330$ to $\langle n \rangle \approx 510$. The bifurcation from single to two steady-states in the MA equations is completely unphysical since (i) moments can only be single-valued, (ii) there is no corresponding transition in the number of maxima of the probability distribution of the CME, and (iii) the two solutions of the MA equations do not provide any meaningful local information on the individual modes of the distribution since the distribution is unimodal (unlike the case for bistable parameter set S1). 

%

%%%%%%%%%%%%%%%%%%%%%%%%%%%%%%
%%%%%%%%%%%%%%%%%%%%%%%%%%%%%%
%%%%%%%%%%%%%%%%%%%%%%%%%%%%%%
\subsection{A deterministic oscillatory system}

Consider next the Brusselator, a well known oscillating chemical system \cite{Prigogine1968,Lefever1988}
\begin{align}\label{brusselator_reactions}
  2 X + Y \xrightarrow{\quad c_1 \quad} 3X,
  \quad X \xrightarrow{\quad c_2 \quad} Y, 
  \quad \varnothing \xrightleftharpoons[c_4]{\quad c_3 \quad} X.
\end{align}
The deterministic rate equations predict sustained oscillations in certain parameter regimes, and damped or no oscillations in other regimes. In contrast, stochastic simulations show that the mean molecule numbers either exhibit damped oscillations or no oscillations \cite{Toner2013}; sustained oscillations are only seen in individual trajectories of the SSA but due to dephasing between independent trajectories, the mean molecule numbers (calculated over an ensemble of trajectories) only show damped oscillations. We now study the MA predictions for this system.  

\begin{figure}[t]
\centering
  \includegraphics[scale=0.47]{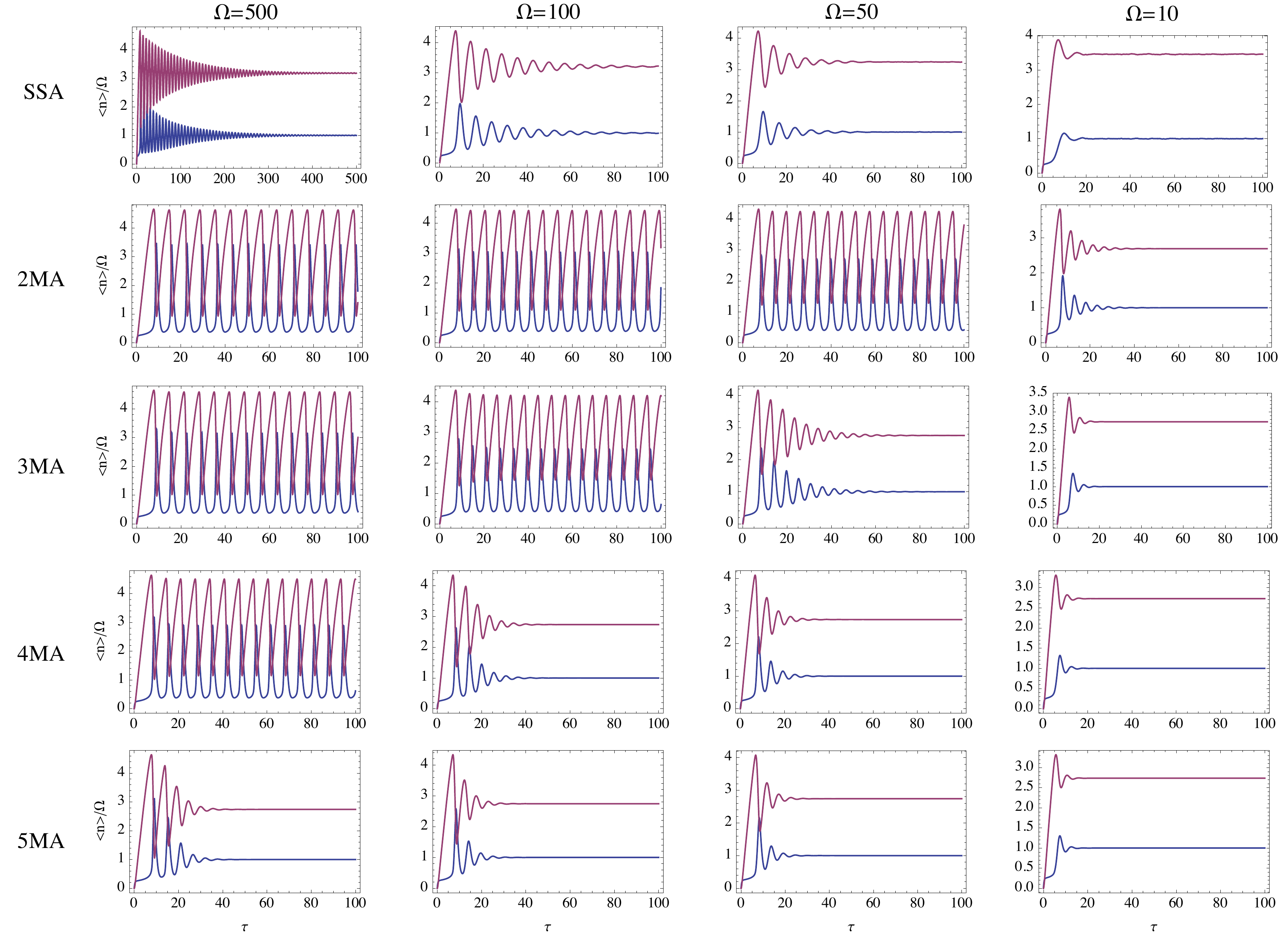}  
  \caption{Comparison of the time evolution of the mean molecule numbers according to the CME and to the MA equations for the Brusselator system \eqref{brusselator_reactions} as a function of the volume $\Omega$. The rate parameters are $c_1 / c_4 = 1, c_2 / c_4 = 3, c_3 / c_4 = 1$. Blue and red lines denote the solutions for species $X$ and $Y$ respectively. Note that the MA equations predict a transition from sustained to damped oscillations as the volume decreases but the CME predicts no such transition; the transition is thus artifactual also since the critical volume at which it occurs increases with the order of the MA.}
  \label{figure_brusselator_time}
\end{figure}

We fix the rate constants to $c_1 / c_4 = 1, c_2 / c_4 = 3, c_3 / c_4 = 1$ and study the MA equation properties as a function of the volume $\Omega$. The deterministic rate equations exhibit sustained oscillations while the mean molecule numbers computed from the SSA show damped oscillations (for all tested volumes) before settling to a non-oscillatory steady-state (see top panel of four graphs in Figure \ref{figure_brusselator_time}). In the same figure, we show the solutions of the 2-5MA as a function of the volume. Note that the MA solutions show a transition from sustained oscillations to damped oscillations at a certain critical volume; the MA solution is an approximation for the moments, i.e., ensemble-averaged behaviour of the CME, however no such transition is seen in the CME solution and hence the sustained oscillation solution of the MA equations is to be treated as an artifact. A further proof of the artifactual nature of the transition is that the critical volume at which it occurs increases rapidly with the order of the MA; these critical volumes are $\Omega = 22, 89, 320, 4900$ for the 2-5MA equations (denoted collectively as $\Omega_2$). This picture is consistent with the critical volume going to infinity in the limit of infinite order of the MA which would imply agreement with the ensemble-average results of the CME in this limit. As for the unphysical bifurcation in the Schl\"ogl system, the origin of the artificial transition from sustained to damped oscillations is unclear since it does not appear to be related to a sudden breakdown of the cumulant neglect assumption at the heart of the MA equations.

\begin{figure}[t]
\centering
  \includegraphics[scale=0.6]{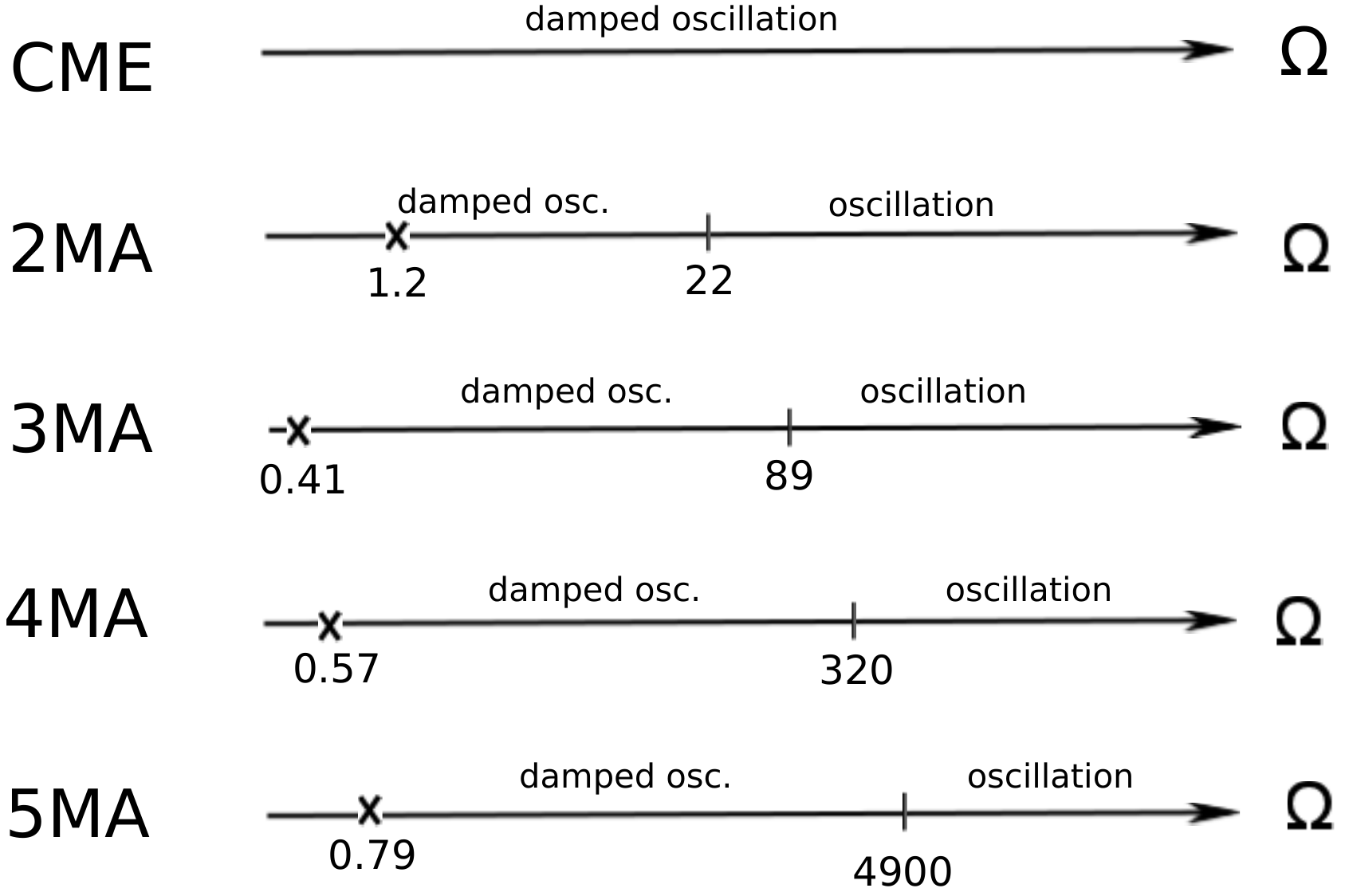}  
  \caption{Damped and sustained oscillation regimes of the MA solution as a function of the volume $\Omega$ for the Brusselator reaction scheme \eqref{brusselator_reactions} with rate constants $c_1 / c_4 = 1, c_2 / c_4 = 3, c_3 / c_4 = 1$. The crosses denote the critical volumes above which all time trajectories for the mean and even central moments are positive and convergent to a steady-state.}
  \label{figure_brusselator_arrows}
\end{figure}

As for previous systems, another critical volume is found below which the trajectories of the MA equations do not converge or lead to unphysical moments. These critical volumes are found to be $\Omega = 1.4, 0.41, 0.57$ and $0.79$ for the 2-5MAs (denoted collectively as $\Omega_1$). These correspond to a range of mean molecule numbers of species $X$ between $0.4$ and $1.4$ and for species $Y$ between $8$ to $9$. A schematic summary of our numerical analysis, including the transition behaviour earlier discussed, is shown in Fig. \ref{figure_brusselator_arrows}. 

The time trajectory and transition analysis put together lead us to the conclusion that the MA equations for the Brusselator lead to physically meaningful predictions only in the finite range of volumes $(\Omega_1,\Omega_2)$ (and the associated finite range of molecule numbers); this conclusion is similar to that for the bistable parameter set S1 of the Schl\"ogl system in Section III B. The analysis here was specifically done for the rate constants $c_1 / c_4 = 1, c_2 / c_4 = 3, c_3 / c_4 = 1$; we have verified that the same picture and conclusions emerge from studying twelve other sets of rate constants (see Table II). Note that in 8 cases, there is no regime where the steady-state solution of the 2MA equations is physically meaningful since the solution is oscillatory for all volumes; in contrast there is just one case where the 3MA suffers a similar problem and no such cases are found using the 4 and 5MA. These results suggests that our analysis broadly holds for all parameters such that the deterministic rate equations predict sustained oscillations. 

\begin{table}
\begin{center}
  \begin{tabular}{|c|c|c|c|c|c|c|c|c|c|c|}
  \hline
  \multicolumn{3}{|c} {parameters} &  \multicolumn{2}{|c} {2MA} &  \multicolumn{2}{|c} {3MA} &  \multicolumn{2}{|c} {4MA} &  \multicolumn{2}{|c|} {5MA} \\ \hline
  $c_1/c_4$ & $c_2/c_4$ & $c_3/c_4$ & $~\Omega_1~$ & $~\Omega_2~$ & $~\Omega_1~$ & $~\Omega_2~$ & $~\Omega_1~$ & $~\Omega_2~$ & $~\Omega_1~$ & $~\Omega_2~$  \\ \hline \hline
  $ \frac{1}{2} $ & $ 2 $ & $ \frac{1}{2} $ & $ 1.8 $ & $ 1.8 $ & $ 0.67 $ & $ 5.7 $ & $ 0.52 $ & $ 17 $ & $ 0.52 $ & $ 44 $ \\ \hline $
 \frac{1}{2} $ & $ 2 $ & $ 1 $ & $ 0.9 $ & $ 14 $ & $ 0.37 $ & $ 350 $ & $ 0.44 $ & $ 4900 $ & $ 0.67 $ & $ ~> 10^6 $ \\ \hline $
 \frac{1}{2} $ & $ 4 $ & $ \frac{1}{2} $ & $ 3 $ & $ 3 $ & $ 1.1 $ & $ 1.1 $ & $ 0.4 $ & $ 4.6 $ & $ 0.38 $ & $ 8 $ \\ \hline $
 \frac{1}{2} $ & $ 4 $ & $ 1 $ & $ 1.5 $ & $ 1.5 $ & $ 0.56 $ & $ 5.9 $ & $ 0.4 $ & $ 12 $ & $ 0.61 $ & $ 21 $ \\ \hline $
 \frac{1}{2} $ & $ 4 $ & $ 2 $ & $ 0.77 $ & $ 28 $ & $ 0.23 $ & $ 88 $ & $ 0.34 $ & $ 210 $ & $ 0.38 $ & $ 410 $ \\ \hline $
 1 $ & $ 2 $ & $ \frac{1}{2} $ & $ 1.8 $ & $ 1.8 $ & $ 0.67 $ & $ 23 $ & $ 0.67 $ & $ 103 $ & $ 0.65 $ & $ ~> 10^6 $ \\ \hline $
 1 $ & $ 4 $ & $ \frac{1}{2} $ & $ 1.8 $ & $ 1.8 $ & $ 1.1 $ & $ 3.9 $ & $ 0.57 $ & $ 10 $ & $ 0.9 $ & $ 17 $ \\ \hline $
 1 $ & $ 4 $ & $ 1 $ & $ 1.5 $ & $ 1.5 $ & $ 0.56 $ & $ 20 $ & $ 0.52 $ & $ 40 $ & $ 0.79 $ & $ ~> 10^6 $ \\ \hline $
 2 $ & $ 2 $ & $ \frac{1}{2} $ & $ 1.8 $ & $ 27 $ & $ 0.72 $ & $ 710 $ & $ 0.88 $ & $ 9700 $ & $ 1.3 $ & $ ~> 10^6 $ \\ \hline $
 2 $ & $ 4 $ & $ \frac{1}{2} $ & $ 1.5 $ & $ 1.5 $ & $ 1.1 $ & $ 11 $ & $ 0.79 $ & $ 25 $ & $ 1.2 $ & $ 42 $ \\ \hline $
 2 $ & $ 4 $ & $ 1 $ & $ 1.5 $ & $ 56 $ & $ 0.59 $ & $ 170 $ & $ 0.54 $ & $ 420 $ & $ 0.77 $ & $ 830 $ \\ \hline $
 4 $ & $ 4 $ & $ \frac{1}{2} $ & $ 3 $ & $ 3 $ & $ 1.1 $ & $ 41 $ & $ 1 $ & $ 80 $ & $ 1.6 $ & $ 130 $ \\ \hline 
  \end{tabular}
  \caption{The table shows the two volumes ($\Omega_1$ and $\Omega_2$) between which the MA equations have a physically meaningful solution for the Brusselator reaction scheme in Eq.~\eqref{brusselator_reactions}. The data is generated for 12 distinct parameter sets. In the volume range $\Omega_1-\Omega_2$ the time trajectories of the MA are convergent to a unique steady-state and exhibit physically meaningful moments at all times (positive mean and even central moments). Above $\Omega_2$, the moment solutions of the MA exhibit artificial sustained oscillations (the SSA always has damped oscillations) and below $\Omega_1$ the time trajectories do not converge or else there exist negative mean and even central moments at a point in time. We consider all combinations for the parameters $c_1/c_4$, $c_2/c_4$ and $c_3/c_4$ drawn from the set $\{ 1/2,1,  2, 4\}$ for which the deterministic rate equations predict sustained oscillations. The initial condition is zero mean and higher-order moments.}
  \label{table_brusselator}
  \end{center}
\end{table}  

%%%%%%%%%%%%%%%%%%%%%%%%%%%%%%
%%%%%%%%%%%%%%%%%%%%%%%%%%%%%%
%%%%%%%%%%%%%%%%%%%%%%%%%%%%%%
\section{Summary and conclusion}\label{sec_conclusion}

In this paper we have elucidated, by means of several exemplary reaction systems, the range of validity of a popular class of MA equations. In particular our numerical results suggest that the solutions of these equations are only physically meaningful when the steady-state mean molecule numbers obtained from the CME are above a certain threshold for deterministic monostable systems and when these molecule numbers fall within a certain finite range of mean molecule numbers for deterministic bistable and oscillatory systems. Our results have important implications for the use of MA approaches to either predict the stochastic dynamics of chemical systems \cite{Ullah2009,Verghese2007} or for parameter inference \cite{Milner2013,Zechner2012}. 

By physically meaningful solutions we specifically mean that (i) the mean molecule numbers and the even central moments of the fluctuations in molecule numbers predicted by the MA equations are positive real numbers at all times and they converge to steady-state values whenever the CME has a stationary solution, (ii) the moments are unique in the sense that the same steady-state moments can be reached from all initial conditions, and (iii) the moments do not exhibit sustained oscillations in the limit of long times. The first two properties are self-evident while the third may not be immediately obvious -- this stems from the fact that though individual trajectories of the SSA can exhibit sustained oscillations, all these noisy trajectories are not in phase since they are independent (provided the rate constants are time independent) and hence an ensemble average leads to damped oscillations through destructive interference between the trajectories. We found that deterministic monostable systems suffer from a breakdown of property (i) for mean molecule numbers below a certain value, deterministic bistable systems suffer from the same and as well from a breakdown of property (ii) for mean molecule numbers above a certain value, while deterministic oscillatory systems have the same property as monostable systems and as well suffer from a breakdown of property (iii) for mean molecule numbers above a certain value. 

We have shown that the transition in the number of steady-state solutions (from 1 to 2) which lead to a breakdown of property (ii) and the transition from damped to oscillatory long time behaviour which lead to the breakdown of property (iii) occur at increasing higher mean molecule numbers as the order of the MA equations is increased -- this strongly suggests that these transitions are unphysical. This conclusion was further supported by showing that the transitions are uncorrelated with sudden changes in the number of maxima of the probability distribution of the CME or in the moment dynamics of the CME. 

We note that above the mean molecule numbers for which properties (ii) and (iii) breakdown, the solution of the moment equations closely resembles that of the deterministic rate equations rather than the moments of the probability distribution of the CME. However as noted above, the range of mean molecule numbers for which all three properties are satisfied is found to increase rapidly as the order of the MA equations is increased -- this means that the solution of the MA resembles less the deterministic rate equations and more the CME as the order is increased. This result is consistent with the fact that the first-order MA is either the same or approximately equal to the deterministic rate equations (since the 1MA involves setting the variance of fluctuations to zero) while the infinite order MA is equivalent to the exact moments of the CME. 
 
We emphasise that the results here reported are specifically for bursty gene expression and a genetic negative feedback loop (monostable systems), the Schl\"ogl model (a bistable system) and the Brusselator (an oscillatory system); it remains to be seen whether the conclusions obtained herein are common to all deterministic monostable, bistable and oscillatory systems, albeit its unlikely that the latter can be conclusively proved since the MA framework is typically only amenable to numerical analysis.  

In conclusion, we have here elucidated the conditions necessary for the validity of the solution of the MA equations. Our results suggest that though this approach presents an efficient computational means of obtaining approximate solutions to the moments of the CME, there are several pitfalls in the interpretation of the results particularly for deterministic bistable and oscillatory systems; these difficulties can be only be overcome through the systematic comparison of the solutions of the MA equations of a certain order with those of higher orders.

\section*{Acknowledgments}

G.S. acknowledges support from the European Research Council under grant MLCS 306999.

%%%%%%%%%%%%%%%%%%%%%%%%%%%%%%
%%%%%%%%%%%%%%%%%%%%%%%%%%%%%%
%%%%%%%%%%%%%%%%%%%%%%%%%%%%%%
%%%%%%%%%%%%%%%%%%%%%%%%%%%%%%
\begin{appendix}

%%%%%%%%%%%%%%%%%%%%%%%%%%%%%%
%%%%%%%%%%%%%%%%%%%%%%%%%%%%%%
%%%%%%%%%%%%%%%%%%%%%%%%%%%%%%
\section{Origin of the unphysical bifurcation in the MA equations}

\begin{figure}[t]
\centering
  \includegraphics[scale=0.44]{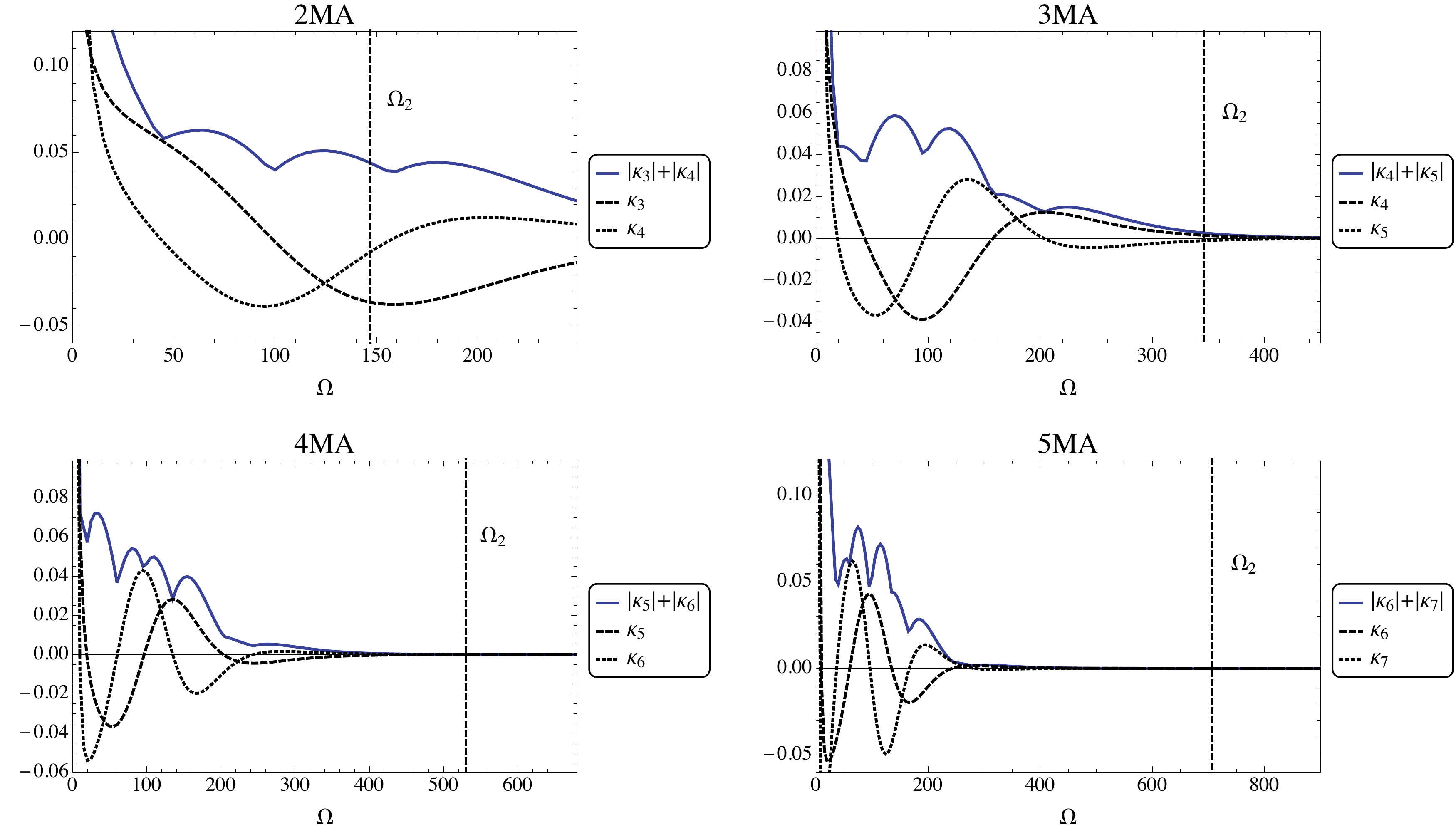}  
  \caption{Cumulants of the concentrations for the Schl\"ogl reaction scheme in \eqref{schoegl_reactions} with parameter set S1, as a function of the volume $\Omega$. $\kappa_n$ denotes the cumulant of order $n$ which is computed from the exact solution of the CME. The solid vertical lines show the critical volumes ($\Omega_2$) at which the MA equations change from having a single to two stable fixed points with positive mean and even central moments. Note that there is no sudden change in the cumulants as a function of volume which anticipates the transition observed in the solution of the MA equations.}
  \label{figure_schloegl_cumulants}
\end{figure}

One is naturally led to the question if there is a way to predict the unphysical bifurcation in the MA equations for the Schl\"ogl model studied in Section III B. Recall that the derivation of the MA equations requires that some of the cumulants are zero. This is of course an assumption and hence one could surmise that perhaps this assumption breaks down dramatically when the MA equations experience the unphysical bifurcation, e.g., the mentioned cumulants of the probability distribution solution of the CME may suddenly take large values as the volume crosses a certain threshold. We now test this hypothesis. Since the reaction scheme \eqref{schoegl_reactions} includes bimolecular and trimolecular reactions, to get a closed set of equations for the first $N$ moments we have to set the $(N+1)$th and $(N+2)$th cumulants to zero. We denote the $N$th cumulant by $\kappa_N$ in the following. We compute the exact values of these two cumulants from the stationary solution of the CME and in particular plot their dependence with the volume. The results are shown in Figure \ref{figure_schloegl_cumulants}. We find that the behaviour of the cumulants gives no apparent hint to the bifurcation of the MAs from one to two positive stable fixed points. Thus our present analysis is inconclusive as to origin of the unphysical bifurcation in the MA equation solutions. 

\section{Information contained in the bistable solution of the 2MA equations}

Since the bistable solution of the MA equations cannot be interpreted as an approximation to the moment of the probability distribution of the CME, the question arises if they can be meaningfully interpreted otherwise. In the bistable regime which appears when solving the 2MA equations in small volumes, the solution of the CME is unimodal, and there is no interpretation of the bistability. For large volumes, however, it turns out that the moments of the two 2MA fixed points provide a good approximation to the moments of the individual peaks of the bimodal probability distribution of the CME. This can be seen by constructing probability distribution functions from the MAs in the bistable regime in the following way. First, we construct two Gaussians with the respective mean and variance obtained from the two solutions of the 2MA. Secondly a linear superposition of these two Gaussians is created; the two superposition weights are calculated from the CME probability distribution by numerically calculating the area under each peak. 

\begin{figure}[t]
\centering
  \includegraphics[scale=0.45]{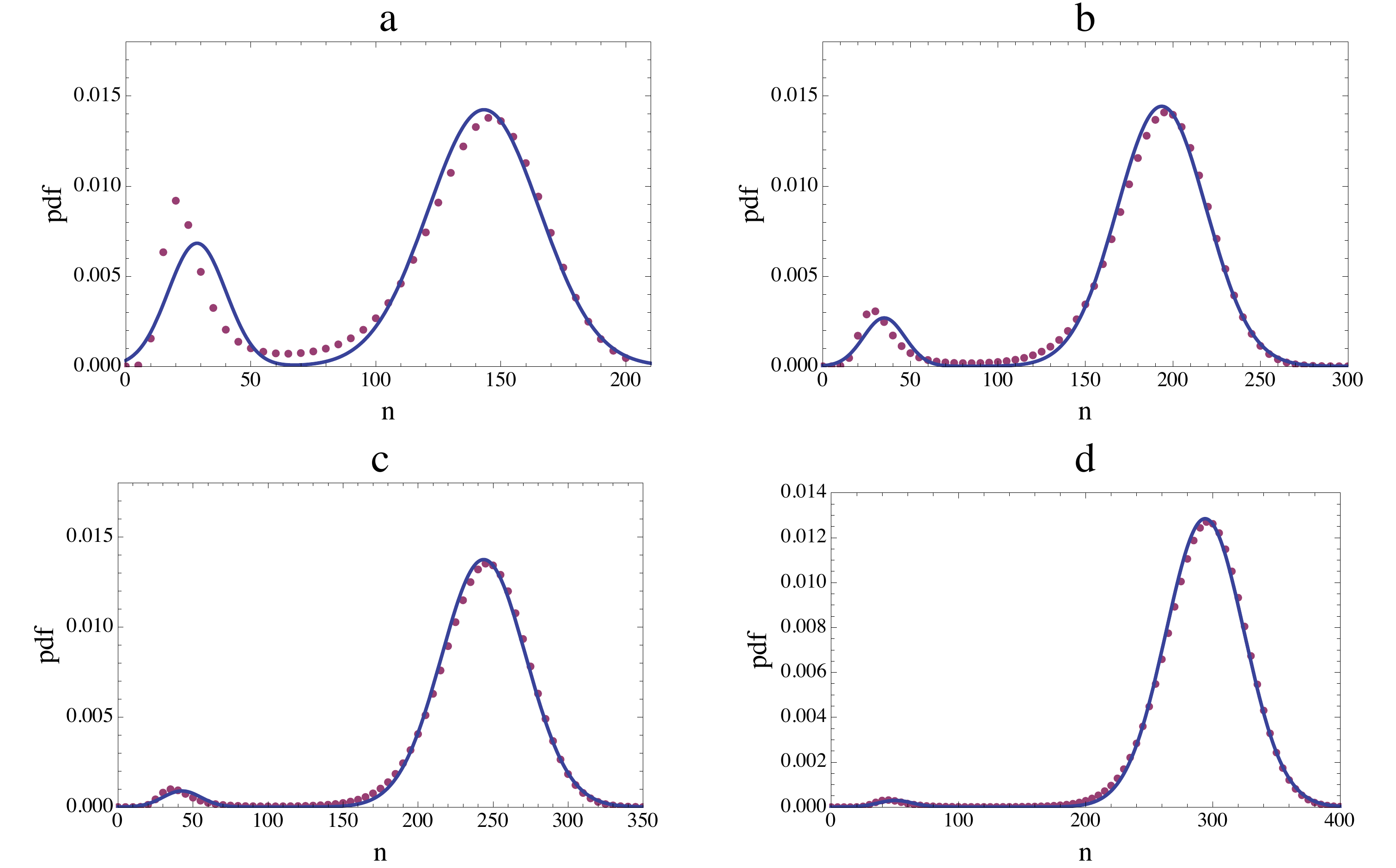}  
  \caption{A comparison of the exact probability distribution function (pdf) obtained from CME (magenta dots) and the reconstructed pdf from the 2MA equations (solid blue line) for the Schl\"ogl reaction scheme in \eqref{schoegl_reactions} with parameter set S1. The volumes are $\Omega = 150, 200, 250, 300$ for (a) - (d) respectively. The 2MA pdf is constructed from the two steady-state solution of the 2MA as detailed in Appendix B.}
  \label{figure_schloegl_pdfs}
\end{figure}

Figure \ref{figure_schloegl_pdfs} shows the exact CME solution (magenta dots) and the probability distribution constructed as detailed above from the 2MA (blue curve) for four different volumes ($\Omega = 150, 200, 250, 300$). There is reasonably good agreement between the two probability distribution functions. However, note that to our knowledge there is no method to obtain the weights self-consistently from the MA equations (we estimated these from the CME itself) and hence all one can conclude is that \emph{in the large volume bistable regime, the 2MA provides information on the individual peaks of the bimodal CME distribution but not on the whole distribution itself. It is also the case that there are volumes for which the CME distribution is bimodal but for which the 2MA equations are monostable ($30 < \Omega < 147$), in which case no reconstruction of the probability distribution function by the aforementioned method is possible.} 

\end{appendix}
\end{document}